\documentclass[aps,prl,,twocolumn,showpacs,superscriptaddress]{revtex4-1}

\usepackage{graphicx} 
\usepackage{bm}  
\usepackage{color,soul}  
\usepackage{ulem}
\usepackage{amssymb}
\usepackage{natbib}

\newcommand{\bra}{\langle}
\newcommand{\ket}{\rangle}
\newcommand{\rcite}[1]{Ref.~\citenum{#1}}
\newcommand{\eref}[1]{Eq.~\ref{#1}}

\newcommand{\fref}[1]{Fig.~\ref{#1}}

\newcommand{\cref}[1]{Chp.~\ref{#1}}

\newcommand{\gae}{\lower 2pt \hbox{$\,
\buildrel{\scriptstyle >}\over {\scriptstyle \sim}\,$}}
\newcommand{\lae}{\lower 2pt \hbox{$\,
\buildrel{\scriptstyle <}\over {\scriptstyle \sim}\,$}}

\begin{document} 

\title{Measuring a topological transition in an artificial spin 1/2 system} 

\author{M. D. Schroer}
\affiliation{JILA, National Institute of Standards and Technology and the University of Colorado, Boulder, Colorado 80309, USA}
\author{M. H. Kolodrubetz}
\affiliation{Physics Department, Boston University, Boston, MA 02215, USA}
\author{W. F. Kindel}
\affiliation{JILA, Department of Physics, University of Colorado, Boulder, CO 80309, USA}
\author{M. Sandberg}
\author{J. Gao}
\author{M. R. Vissers}
\author{D. P. Pappas}
\affiliation{National Institute of Standards and Technology, Boulder, CO 80305, USA.}
\author{Anatoli Polkovnikov}
\affiliation{Physics Department, Boston University, Boston, MA 02215, USA}
\author{K. W. Lehnert}
\affiliation{JILA, National Institute of Standards and Technology and the University of Colorado, Boulder, Colorado 80309, USA}
\affiliation{Department of Physics, University of Colorado, Boulder, CO 80309, USA}

\date{\today}


\begin{abstract}
We present measurements of a topological property, the Chern number ($C_\mathrm{1}$), of a closed manifold in the space of two-level system Hamiltonians, where the two-level system is formed from a superconducting qubit. We manipulate the parameters of the Hamiltonian of the superconducting qubit along paths in the manifold and extract $C_\mathrm{1}$ from the nonadiabitic response of the qubit. By adjusting the manifold such that a degeneracy in the Hamiltonian passes from inside to outside the manifold, we observe a topological transition $C_\mathrm{1} = 1 \rightarrow 0$. Our measurement of $C_\mathrm{1}$ is quantized to within 2 percent on either side of the transition.   
\end{abstract}


\maketitle

The topology of quantum systems has become a topic of great interest due to the discovery of topological insulators in two \cite{Haldane1988_1,Kane2005_1,Bernevig2006_1,Konig2007_1} and three \cite{Fu2007_1,Fu2007_2,Hsieh2008_1} dimensions.  It has been shown that certain robust topological invariants, such as the Chern number, help to classify physical phenomena \cite{Thouless1982_1,Niu1985_1,Hatsugai1993_1}.  As these topological integers remain unchanged by small perturbations to the system, jumps in their value represent nontrivial topological transitions in the quantum system, such as an increment in the filling factor of the integer quantum Hall state\cite{Thouless1982_1}. To understand the meaning of the Chern number, consider initializing a system in the ground state of a Hamiltonian described by two parameters, for example the angles $\theta$ and $\phi$ of a magnetic field applied to a spin system.   Adiabatically adjusting these parameters around a closed path that bounds a surface $\mathcal{S}$, one might expect to arrive back at the original ground state, up to a dynamical phase.  However, Berry and Pancharatnam \cite{Berry1984_1,Pancharatnam1956_1} showed that there is an additional phase contribution known as the geometric or Berry's phase ($\varphi_\mathrm{Berry}$).  This phase is given by the surface integral

\begin{equation}
\varphi_\mathrm{Berry} = \int_\mathcal{S} {\bf F} \cdot d{\bf S} ~,
\label{eq:bphase_surfaceint}
\end{equation}
where ${\bf S}$ is a vector normal to the surface and ${\bf F}$ is a vector known as the Berry curvature that characterizes how the ground state is modified by changing parameters\cite{stokes_note}.

If $\mathcal S$ is a closed manifold, then its (non-existent) boundary clearly gives $\varphi_\mathrm{Berry} = 0$.  However, phase is only well-defined up to multiple of $2\pi$ and, although the Berry phase depends on the $U(1)$ gauge choice $|\psi_0\ket \to e^{i\varphi(\theta,\phi)}|\psi_0\ket$ where $|\psi_0(\theta, \phi)\ket$ is the ground state, the Berry curvature is gauge-invariant.  Therefore, the integral
\begin{equation}
C_\mathrm{1} = \frac{1}{2\pi} \oint_\mathcal{S} {\bf F} \cdot {d\bf S}
\label{eq:chintegral}
\end{equation}
is a well-defined topological invariant known as the (first) Chern number\cite{Chern1946_1}, which is quantized to integer values. This Chern number may be intuitively understood as counting the number of times an eigenstate wraps around a manifold in Hilbert space, and is precisely the topological invariant that yields, for example, quantization of the resistance in the integer quantum Hall effect \cite{Thouless1982_1,Niu1985_1,Hatsugai1993_1}.

The Berry phase has been investigated in a wide variety of systems\cite{Tomita1986_1,Zhang2005_1,Novoselov2006_1,Atala2013_1}, both as a fundamental property of quantum systems and as a practical method to manipulate quantum information\cite{Sorensen2000_1,Kerm2013}. In a breakthrough experiment, Leek et al. first measured the Berry phase of a superconducting qubit\cite{Leek2007_1, Berger2012_1,Berger2013_1}.  These experiments typically extract $\varphi_\mathrm{Berry}$ interferometrically after a cycle of closed loop evolution, using a spin echo pulse to remove the dynamical phase.  Such interference experiments do not easily generalize to more complicated Hamiltonians, such as interacting many-body systems. For instance, \cite{Leek2007_1} uses the fact that the Berry phase is identical for the ground and excited states, which will not hold for more than a single spin.

\begin{figure}[ht]
\includegraphics[width= 0.9\linewidth]{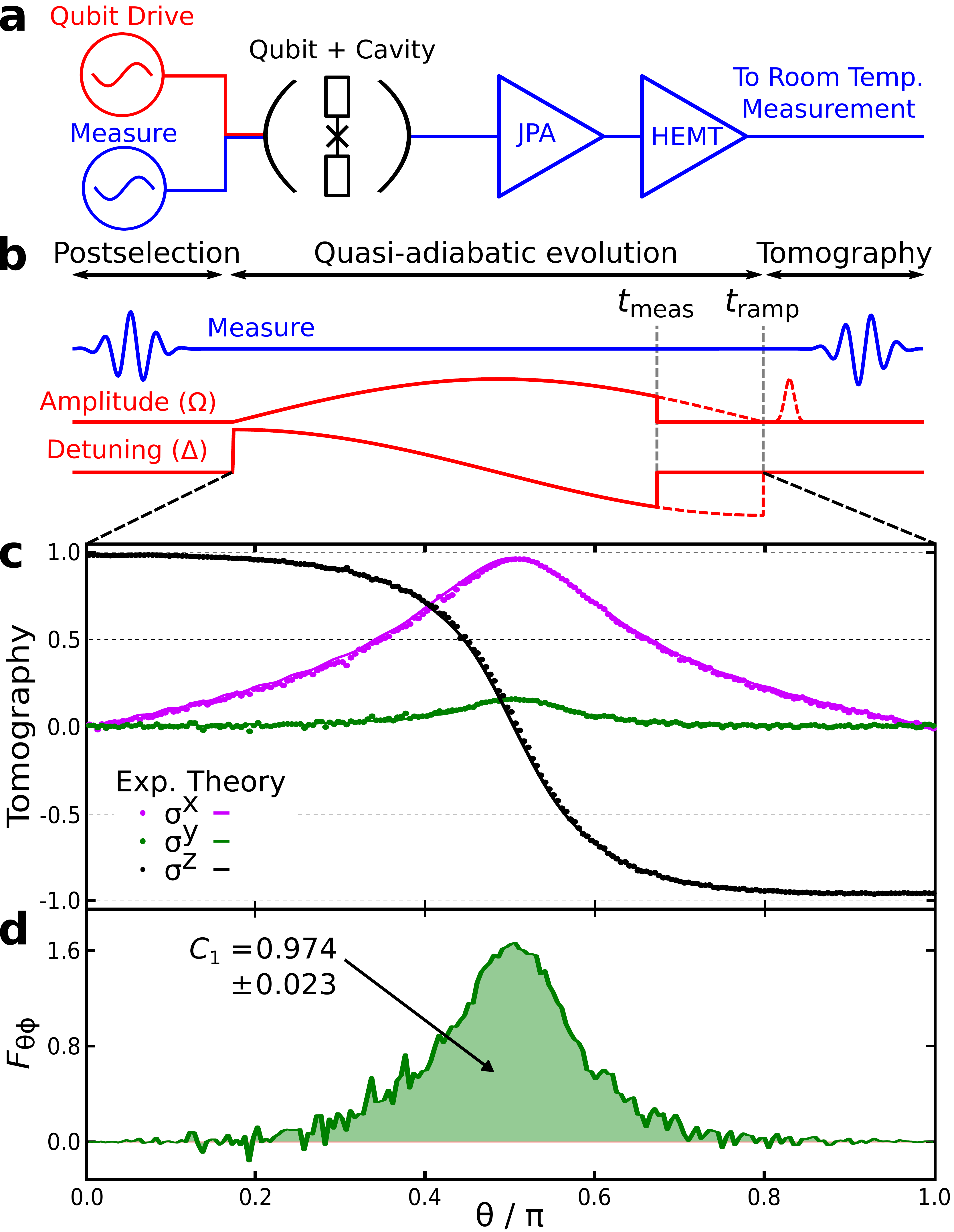}
\centering
\caption{Measuring Berry curvature and Chern number in a transmon qubit.
(a) Diagram of the experimental setup. The qubit is manipulated and probed by two separate microwave generators, while a Josephson parametric amplifier (JPA) provides high fidelity measurement.  For a more complete diagram see the Supplementary Information. (b) Experimental pulse sequence.  Following an initial measurement to project  into the ground state, the detuning and Rabi drive are ramped along an elliptical protocol, after which quantum state tomography is performed. (c)  Tomography of such a ramp, with parameters $t_\mathrm{ramp}= 1~\mu$s, $\Delta_1 / (2\pi) =  30$ MHz, and $\Omega_1 / (2 \pi) = 10$ MHz. Data are shown as solid circles, while the solid lines are a theoretical expectation with $T_1=22~\mu$s and $T_2^*=9~\mu$s, with $\Delta_2/(2\pi)=300$ kHz added to account for uncertainty in the qubit frequency. (d) Using \eref{eq:F_th_phi_expt}, one can extract the Berry curvature $F_{\theta \phi}$, the integral of which gives the Chern number ($C_\mathrm{1}$, see \eref{eq:chernnumber}).}
\label{fig:tomography}
\end{figure}

In this letter, we demonstrate a method of probing topology which is applicable to systems of any size or complexity. As a test of this method, we extract the Chern number of closed manifolds in the parameter space of two level system (qubit) Hamiltonians.  A single qubit has a simple topological structure that may readily be described analytically. However, the experimental method we use to reveal this structure can be applied to more complex systems, such as large spin chains with arbitrary couplings, where simulation on classical hardware is inefficient\cite{Koonin1998_1}.  Following the proposal in \rcite{Gritsev2012_1}, we measure the response of the qubit to nonadiabatic manipulations of its Hamiltonian $H(\theta, \phi)$, which lead to an apparent force ${f_\phi} \equiv -\partial_\phi H$, given by \cite{Berry1989_1, Avron2011_1, Gritsev2012_1} 

\begin{equation}
\bra f_\phi \ket = \bra \psi_0 | f_\phi | \psi_0 \ket - v_\theta F_{\theta \phi} + \mathcal{O}(v^2) ~
\label{eq:gen_force_apt}
\end{equation}
where $v_\theta$ is the rate of change for the parameter $\theta$ and $F_{\theta \phi}$ is a component of the Berry curvature tensor \cite{curvaturetensor_note}.  If the system parameters are adjusted (ramped) slowly enough such that the $\mathcal{O}(v^2)$ terms are negligible, then the Berry curvature may be extracted from this linear response.  By integrating the Berry curvature over a closed parameter manifold, we extract $C_\mathrm{1}$.  When the manifold encloses a single degeneracy in the Hamiltonian, we find $C_\mathrm{1} \approx 1$, and when it encloses no degenerate points we find $C_\mathrm{1} \approx 0$, thus we observe a topological transition in a spin 1/2 system.  The transition from $C_\mathrm{1} = 1$ to $C_\mathrm{1} = 0$ is accurately quantized to within 2\%.  The simplicity and generality of this method makes it an attractive means of probing the topology of engineered quantum systems.

In our experiment we use a transmon qubit primarily made of Titanium Nitride, as described in \rcite{sandberg2013_1}. The qubit is operated in the strong dispersive circuit QED regime\cite{Koch2007_1}, and cooled in an aluminum 3D microwave cavity using a dilution refrigerator with a base temperature below 25 mK. A Josephson Parametric Amplifier (JPA) was used in phase sensitive mode to perform high-fidelity single-shot readout of the qubit\cite{Castellanos-Beltran2008_1,vijay2011_1,Riste2012,lin2013_2} from the qubit state-dependent phase shift of a probe tone\cite{Walraff2005_1}.  Starting from a mixed state with $\sim 5$\% excited state population, the qubit was initialized in its ground state with $\sim 98.8$\% fidelity by  strongly measuring the state of the qubit and post-selecting data from initially-measured ground states.  Fig.~\ref{fig:tomography}a depicts a simplified system schematic; see the Supplementary Materials for more details.

The transmon is effectively a non-linear LC resonator \cite{Koch2007_1}, with a transition frequency of $\omega_q = 4.395$ GHz.  An anharmonicity of 280 MHz makes the qubit an effective two level system in the parameter regimes explored here.    In the rotating frame of an applied microwave drive of frequency $\omega_{d}$, the Hamiltonian for the qubit may be written as\cite{CohenTannoudji1993_1,Koch2007_1}
\begin{equation}
\boldsymbol{H}/\hbar=\frac{1}{2}\left[ \Delta ~ \boldsymbol{\sigma_z} + \Omega ~\boldsymbol{\sigma_x} \cos \phi + \Omega ~\boldsymbol{\sigma_y} \sin \phi \right]~,
\label{eq:H_rwa}
\end{equation}
where $\Delta = \omega_d - \omega_q$, $\phi$ is the phase of the drive tone, $\Omega$ expresses the amplitude of the drive tone as the Rabi oscillation frequency it would induce if it were applied at the qubit resonance, and $\boldsymbol{\sigma_x}$, $\boldsymbol{\sigma_y}$ and $\boldsymbol{\sigma_z}$ are the Pauli spin matrices.  By varying these parameters, we can create arbitrary single-qubit Hamiltonians.  In particular, we work with collections of Hamiltonians that can be represented in parameter space as an ellipsoidal manifold given by
\begin{equation}
\Delta = \Delta_1 \cos \theta + \Delta_2~,~\Omega=\Omega_1 \sin \theta ~,
\label{eq:H_params}
\end{equation}
with cylindrical symmetry about the $z$-axis.  Throughout this paper, we work with ellipsoids of size $\Delta_1/(2\pi) = 30$ MHz and $\Omega_1/(2\pi) = 10$ MHz, and  vary $\Delta_2$ between -10 and 60 MHz.  These particular parameter manifolds are not a important choice, as the topological properties we extract are expected to be independent of deformations of the manifold that do not cross the degeneracy at $\Delta = \Omega = 0$.

\begin{figure*}[ht]
\centering
\includegraphics[width= 0.9\linewidth]{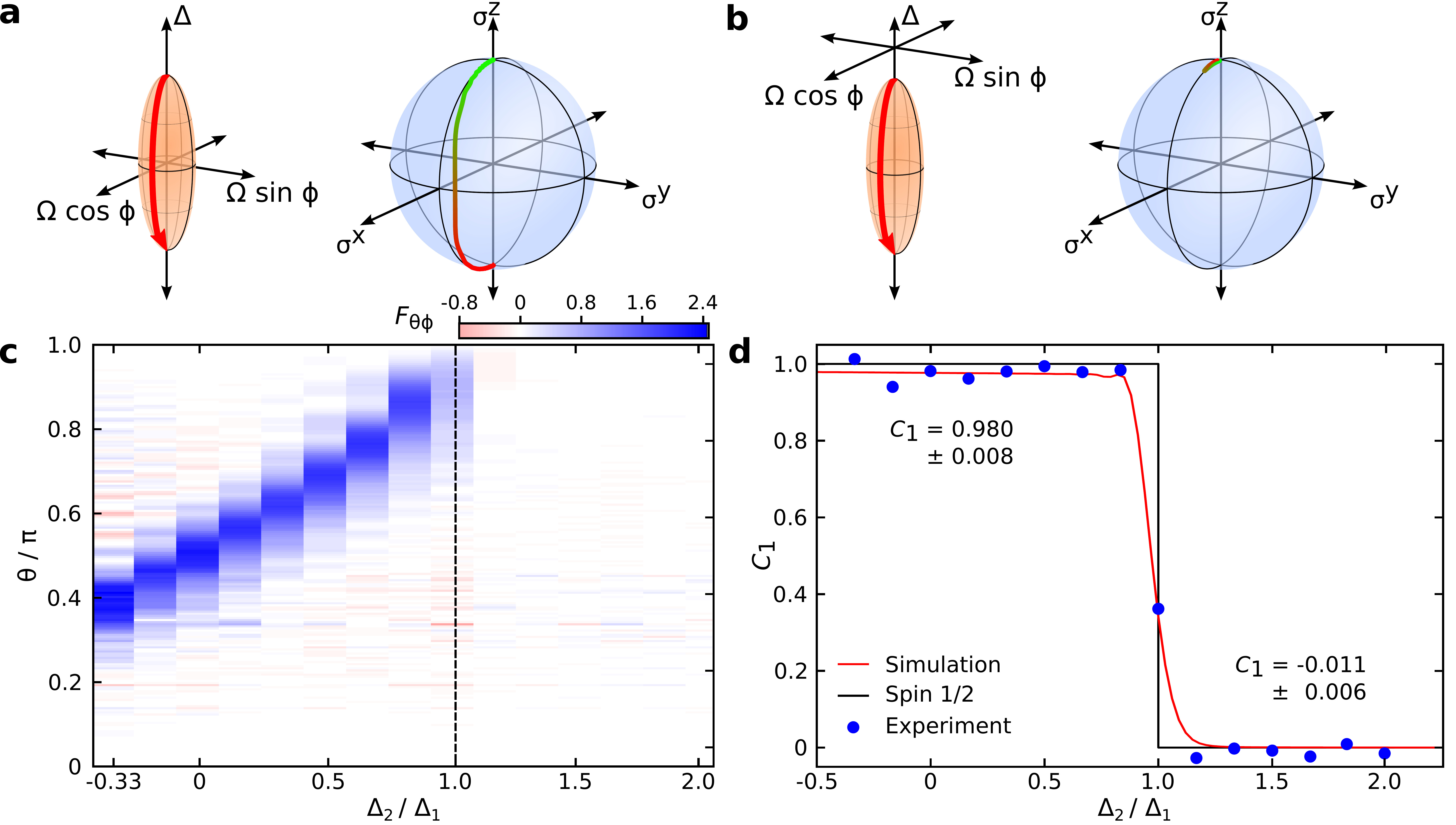}
\caption{Measuring the topological transition.  (a) A schematic of a parameter sweep for $\Delta_2 = 0$.  The orange ellipse represents the parameter surface used to measure $C_\mathrm{1}$, with the particular sweep shown in red.  The results of a simulation of the ramp without relaxation are plotted on the Bloch sphere on the right.   The state wraps the Bloch sphere, with a deviation in the $\sigma^y$ that represents the measured signal.  (b) The same as for (a), with $\Delta_2 = 1.5 \Delta_1$.  Here, there is no wrapping of the Bloch sphere.  (c)  The Berry Curvature measured as a function of $\Delta_2 / \Delta_1$, which is integrated in (d) to yield $C_\mathrm{1}$. }
\label{fig:transition}
\end{figure*}

Figure \ref{fig:tomography}b depicts a typical sequence used to measure the Berry curvature. We first initialize the qubit in its ground state at $\theta(t=0) =0$\cite{rwa_groundstate_note}, fix $\phi(t)=0$ and linearly increase (ramp) the angle $\theta(t)=\pi t / t_\mathrm{ramp} $  in time,  stopping the ramp at various times $t_\mathrm{meas} \leq t_\mathrm{ramp}$ to perform qubit tomography.  At each $t_{meas}$, we measure the generalized force

\begin{equation}
\bra f_\phi \ket = -\bra \partial_\phi H \ket\big|_{\phi=0} = -\frac{\Omega_1 \sin \theta}{2} \bra\boldsymbol{\sigma^y}\ket ~,
\label{eq:gen_force}
\end{equation}
which is zero in the adiabatic limit.  Therefore, the Berry curvature is given as the leading order correction to adiabatic manipulation by

\begin{equation}
F_{\theta \phi} = \frac{\bra \partial_\phi H \ket}{v_\theta} = \frac{\Omega_1 \sin \theta}{2 v_\theta} \bra \boldsymbol{\sigma^y}\ket,
\label{eq:F_th_phi_expt}
\end{equation}
By integrating this expression we obtain the Chern number $C_\mathrm{1} = (2\pi)^{-1} \int_0^\pi d\theta \int_0^{2\pi} d\phi F_{\theta \phi}$.  As the Hamiltonian with parameters in \eref{eq:H_params} is cylindrically symmetric about the $z$-axis, the Berry curvature is a function of $\theta$ alone. Thus the Chern number reduces to 
\begin{equation}
C_\mathrm{1} = \int_0^{\pi} F_{\theta \phi} d \theta~.
\label{eq:chernnumber}
\end{equation}

Figure \ref{fig:tomography}c shows the results of state tomography for a protocol with $t_\mathrm{ramp} = 1~\mu$s and $\Delta_2 = 0$.  The data agrees well with a simulation using a Lindblad master equation model\cite{lind_vs_BR_note}, with dissipation set to the experimentally-measured rates $T_1 = 22~\mu$s and $T_2^* = 9~\mu$s\cite{detuningoffset_note}.    We extract the Berry curvature $F_{\theta \phi}$ (\fref{fig:tomography}d) from the measured values of $\bra \sigma_y \ket$ and integrate it to get a measured Chern number of $C_\mathrm{1}=0.974\pm 0.023$, within a few percent of the quantized value $C_\mathrm{1}=1$ expected from theory.  Correcting for the finite fidelity preparation of the ground state\cite{finitet_correction_note}, we find  $C_\mathrm{1} =  0.998 \pm 0.023$. \cite{Koonin1998_1}

To drive a topological transition in the qubit, we now modify the detuning offset $\Delta_2$. At $\Delta_2=0$, the Chern number of a single qubit is $C_\mathrm{1}=1$, which counts the number of times that the Bloch vectors wrap around the sphere as $\theta$ and $\phi$ are varied. One can see this by examining the limits $\theta=0$ and $\theta=\pi$, which correspond to ground states $\bra \psi_0 | \sigma_z |\psi_0 \ket = 1$ and $-1$ respectively.  Since the wavefunction is opposite at the poles, it must wrap the sphere in between (see \fref{fig:transition}a).

As we change $\Delta_2$, the ground state evolution is quantitatively modified, but for $|\Delta_2| < |\Delta_1|$, the Chern number remains unchanged.  However, for $|\Delta_2| > |\Delta_1|$ the ground state matches at the two poles (see \fref{fig:transition}b). This gives a Chern number of zero, meaning that the system undergoes a topological transition at $|\Delta_2| = |\Delta_1|$.   Such a transition may only occur when the Berry curvature becomes ill defined at the point ${\Delta = \Omega = 0}$. The topological transition corresponds to moving this degeneracy from inside to outside the elliptical manifold.

The measured Chern number is plotted in \fref{fig:transition}d, showing a relatively sharp
transition at the expected value $\Delta_2 = \Delta_1$. Experimentally, the transition is broadened  due to the non-zero ramp rate, which can be understood as the influence of the $O(v^2)$ terms in \eref{eq:gen_force_apt} (see Supp. Inf.). The average values of Chern number for $\Delta_2 < \Delta_1$ [$\Delta_2 > \Delta_1$] are $C_\mathrm{1}=0.980(8)~[-0.011(6)]$, in good agreement with theory even without correction for infidelity in the state preparation. The topological transition should become sharper in the limit $t_\mathrm{ramp} \to \infty$ for a perfectly coherent qubit. However, the finite T$_1$ and T$_2^*$ values for a real qubit broaden this transition and destroy perfect quantization\cite{Xu2014_1,gapless_note} even in the limit of fully adiabatic evolution.  Additionally, due to the linear dependence of the generalized force on the velocity in \eref{eq:gen_force_apt}, the number of measurements required for constant signal to noise scales as  $1/{t_\mathrm{ramp}}^2$.  Given the properties of the qubit used here, we found $t_\mathrm{ramp} = 1~\mu$s  to be a good compromise between these competing behaviors.

\begin{figure}[t]
\centering
\includegraphics[width=\linewidth]{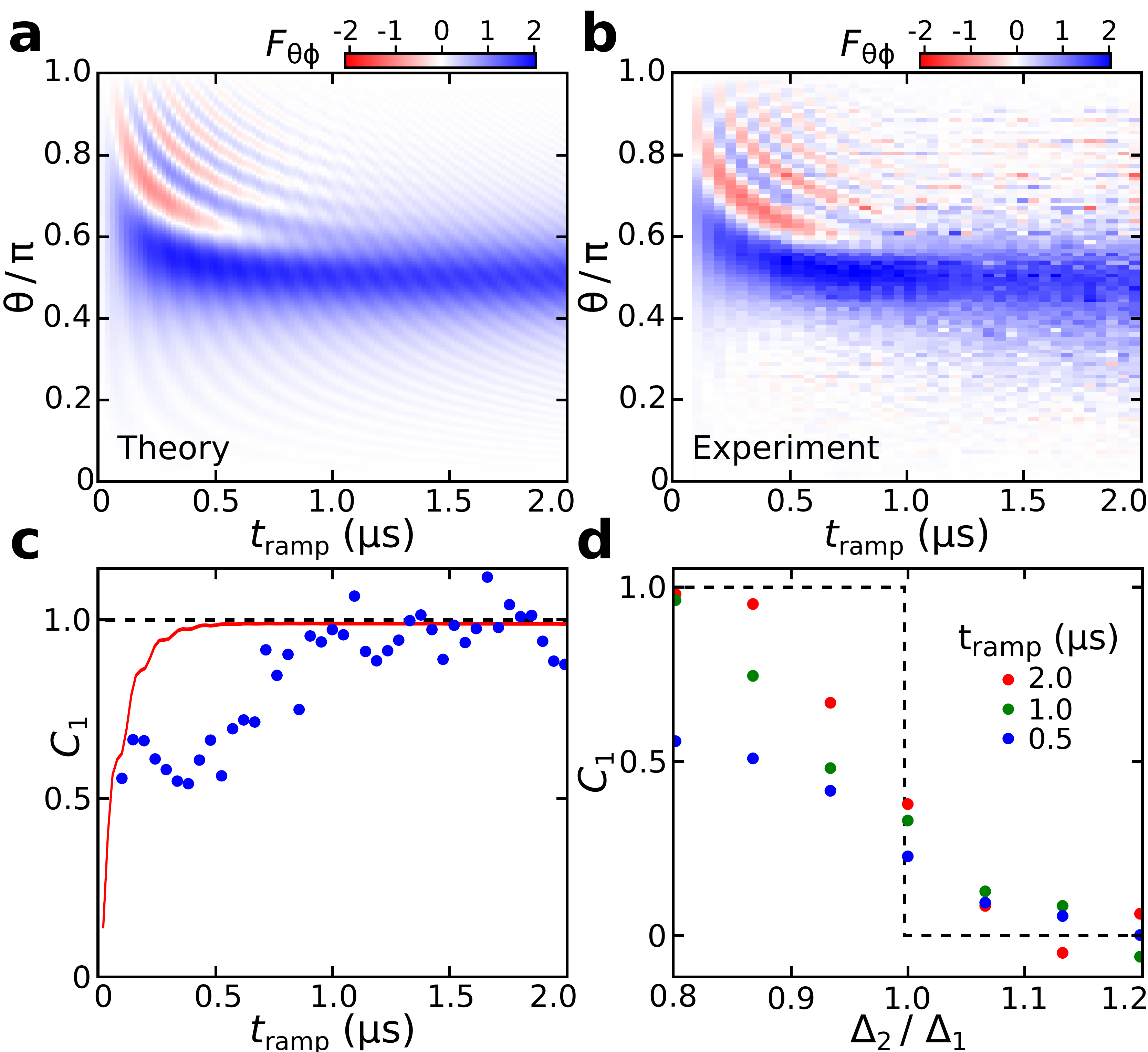}
\caption{The effects of finite velocity. (a-b) The Berry curvature measured as  a function of $t_\mathrm{ramp} \propto 1/v_\theta$, for both a simulation with Lindbladian operators and as measured in the experiment.  Here, $\Delta_2 = 0$, and $\theta(t) = \pi t / t_\mathrm{ramp}$.  For $t_\mathrm{ramp} > 1~\mu$s, the data closely matches the values for a spin 1/2 particle, while at higher velocities the higher order terms of \eref{eq:gen_force_apt} dominate. In (c) the data from (b) has been integrated to produce the measured $C_\mathrm{1}$, with the solid line the results of a simulation. The divergence of the data from theory below $\sim 750$ ns is likely due to imperfections in the microwave tone generation.  (d) The topological transition measured for different values of $t_\mathrm{ramp}$, demonstrating the sharpening of the transition observed at smaller velocities.}
\label{fig:ramprate}
\end{figure} 

We explore the effect of finite ramp time by measuring $C_\mathrm{1}$ and the Berry curvature as a function of $t_\mathrm{ramp}$.  \fref{fig:ramprate}(a) and (b) show a numerical simulation and experimental results for the measured Berry curvature as a function of $t_\mathrm{ramp}$.  The data qualitatively matches the theory, with the measured Berry curvature largely independent of  $t_\mathrm{ramp}$ for times longer than $\sim 1~\mu$s.  For shorter $t_\mathrm{ramp}$, we observe oscillations due to the increasingly nonadiabatic passage throughout the ramp.  While these oscillations affect the measured Berry curvature on a point by point basis, they can be expected to partially average away when computing $C_\mathrm{1}$.  \fref{fig:ramprate}(c) shows the values of $C_\mathrm{1}$ computed from the data in (a) and (b).  Here we see that the simulations predict that $C_\mathrm{1} \approx 1$ for  $t_\mathrm{ramp} \gtrsim 400$ ns.  In contrast,  the experimental results depart from the quantized value at $\sim 750$ ns; we expect that this discrepancy is due to increasing error in the drive modulation occurring  at shorter ramp times. 

Finally, \fref{fig:ramprate}d shows the topological transition measured over a finer range of $\Delta_2$, for $t_\mathrm{ramp} = $ 0.5, 1, and 2~$\mu$s.  Fundamentally, the finite coherence  of the qubit is expected to limit this transition to a minimum width of approximately
\begin{equation}
\delta \left(\frac{\Delta_2}{\Delta_1}\right) \approx \frac{2 \pi}{  \Delta_ 1\mathrm{T}_2^* } = 0.02 ~,
\end{equation}
due to broadening of the qubit resonance.  This is less than the width observed for all values of $t_\mathrm{ramp}$, thus we observe a sharpening of the transition at longer ramp times, consistent with expectations.  

In addition to their relevance to quantum information processing, the measurements described in this letter may be considered as a simulation of a condensed matter system using engineered and tunable quantum resources.  Namely, by mapping states on the Bloch sphere to wave vectors in the first Brillouin zone, one can make an analogy to noninteracting many body condensed matter systems\cite{RoushanInPrep2014_1}. To illustrate this concept, we briefly describe in the supplement how to map the topology of a single qubit onto the Haldane model of graphene\cite{see_suppl_note}, a paradigmatic system exhibiting a topological phase transition\cite{Haldane1988_1}.   The high level of control in superconducting qubits make them potentially flexible platforms for simulating the topology of condensed matter systems.

\begin{acknowledgments}
We would like to acknowledge useful discussions with Maxim Vavilov and Pedram Roushan.  This work was supported by the National Science Foundation under Grant Number 1125844, NSF DMR-0907039, AFOSR FA9550-10-1-0110.  This work was supported in part by the Laboratory for Physical Sciences and also performed in part at the NIST Center for Nanoscale Science and Technology. M.D.S. is supported by NRC.
\end{acknowledgments}

\bibliographystyle{apsrev}
\bibliography{References}

\begin{thebibliography}{46}
\bibitem[{sto()}]{stokes_note}
\bibinfo{note}{Note that this version of Stokes' theorem is only valid for a
  three-dimensional parameter space. In this case, the vector form of ${\bf F}$
  is analogous to the magnetic field in electromagnetism, which is given by the
  off-diagonal components of the electromagnetic tensor $F_{\mu \nu}$ according
  to $\epsilon^{i j k} F_k = -F^{i j}$, where $\epsilon$ is the antisymmetric
  tensor\cite{Jackson1999_1}.}
\bibitem[{det()}]{detuningoffset_note}
\bibinfo{note}{The fit was performed with only one free parameter, an effective
  offset field $\Delta_2$, to account for an unexpected shift in the frequency
  of the qubit, which we attributed to a weak Stark shift.}
\bibitem[{fin()}]{finitet_correction_note}
\bibinfo{note}{This approximation is justified for a two level-system as $\bra
  \mathrm{g} | \sigma^i(t)|\mathrm{g} \ket = -\bra \mathrm{e} | \sigma^i(t) |
  \mathrm{e} \ket$, where $| \mathrm{g} \ket$ and $ | \mathrm{e} \ket$ are
  ground and excited state wave functions.}
\bibitem[{gap()}]{gapless_note}
\bibinfo{note}{We can think of relaxation and decoherence as coming from
  coupling to a finite temperature bath, which has gapless excitations.
  Therefore, in the absence of a gap, the Chern number is not formally
  quantized. This effect, however, is small at low temperature which, for
  example, justifies the accuracy of quantization in integer quantum Hall
  systems.}
\end{thebibliography}

\end{document}